\documentclass[10pt,a4paper,twoside]{article}
\usepackage{amsfonts,psfig,graphicx}
\usepackage{etc9macro}

\newcommand{\pr}{\partial}

\newcommand{\be}{\begin{equation}}
\newcommand{\en}{\end{equation}}

\newcommand{\x}{{\bf{x}}}
\newcommand{\n}{{\bf{n}}}

\begin{document}

\begin{etcpaper}

\begin{etctitle}
  \etcsettitle{Buoyant turbulent mixing in shear layers}
              {Buoyant turbulent mixing in shear layers}
  \etcaddauthor{}{Geurts}{B.J.}
  \etcaddaffiliation{}{Faculty of Mathematical Sciences, University of Twente,\break
                        P.O. Box 217, 7500 AE Enschede, The Netherlands}
  \etcemailaddress{b.j.geurts@math.utwente.nl}
  \etcsetpaperid{304}
\end{etctitle}

\section{Introduction}

Buoyancy effects in unstably stratified mixing layers express themselves through gravity currents of heavy fluid which propagate in an
ambient lighter fluid. These currents are encountered in numerous geophysical flows, industrial safety and environmental protection issues
\cite{haertel}. During transition to turbulence a strong distortion of the separating interface between regions containing `heavy' or
`light' fluid arises. The complexity of this interface will be used to monitor the progress of the mixing. We concentrate on the
enhancement of surface-area and `surface-wrinkling' of the separating interface as a result of gravity-effects.

Mixing in a turbulent flow may be visualized by a number of snapshots, e.g., in an animation. However, for a quantitative analysis, more
specific measures are needed. Here we represent the separating interfaces as level-sets of a scalar field $c$ ($0 \leq c \leq 1$). We
establish a close correspondence between the growth of surface-area and wrinkling, in particular at large buoyancy effects. We also show
that this process can be simulated quite accurately using large-eddy simulation with dynamic subgrid modeling. However, the subgrid
resolution, defined as the ratio between filter-width $\Delta$ and grid-spacing $h$, should be sufficiently high to avoid contamination due
to spatial discretization error effects.

\section{Surface-area and wrinkling of level-sets}

To specify the intricate structure of the separating interface we consider global properties of the level-set
defined by $S(t)=\{ \x~|~c(\x,t)=1/2 \}$. The mixing-efficiency $\eta$ is defined in terms of the surface-area $A$ as:
\begin{equation}
\eta(t)=\frac{A(t)}{A(0)}~~~;~~A(t)=\int_{S(t)}dS~=~\int_{V}d\x~\delta(c(\x,t)-\frac12) | \nabla c(\x,t)|
\label{eta}
\end{equation}
Instead of integrating directly over the complex and possibly fragmented interface $S(t)$, we evaluate the volume integral in~\ref{eta}, in
which $S \subset V$, using a locally exact quadrature method developed in {\mbox{\cite{jot2001}.}} A further characterization of the
geometry is obtained by considering the average curvature $C$ and wrinkling $W$:
\begin{equation}
C(t)=\int_{S(t)} dS~ \kappa(\x,t)~~~;~~W(t)=\int_{S(t)} dS~ |\kappa(\x,t)|
\label{candw}
\end{equation}
where the local curvature $\kappa$ is defined as
\begin{equation}
\kappa(\x,t)=\nabla \cdot \n ~~~;~~\n (\x)=\frac{\nabla c(\x)}{|\nabla c(\x)|}
\end{equation}

The effect of gravity is represented by source terms in the momentum and energy equations~\cite{jot2001}. Within the
Boussinesq approximation we have:
\begin{eqnarray}
   &&  
\pr_t\rho +\pr_j(\rho u_j)= 0 \nonumber \\
   &&  
\pr_t(\rho u_i) +\pr_j(\rho u_i u_j)+\pr_i p -\pr_j \sigma_{ij} =-
\Big( \frac{\alpha \rho c}{Fr^{2}} \Big) \delta_{i2} \nonumber \\
   &&
\pr_t e + \pr_j((e+p)u_j) - \pr_j(\sigma_{ij}u_i)+\pr_j q_j =-
\Big( \frac{\alpha \rho c }{Fr^{2}} \Big) u_{2}
\label{nseqs}
\end{eqnarray}
where $\rho$ denotes the `light' fluid density, $u_{i}$ is the velocity in the $x_{i}$ direction, $p$ the pressure, $e$ the total energy
density, $q_{j}$ the heat-flux and $\sigma_{ij}=S_{ij}/Re$ with $S_{ij}$ the rate of strain tensor and $Re$ the Reynolds number. The source
terms on the right hand side represent gravity effects in the negative vertical (i.e. $x_{2}$) direction in which
$\alpha=(\rho_{h}-\rho)/ \rho$ with $\rho_{h}$ the density of the `heavy' fluid. The Froude number $Fr=U^{*}/\sqrt{g^{*}L^{*}}$ with
$g^{*}$ the gravitational constant and $U^{*}$, $L^{*}$ a reference velocity and length scale respectively. We consider $\alpha=1$.
Finally, $c$ denotes a scalar field which relates to the local density ${\tilde{\rho}}$ as ${\tilde{\rho}}=(1-c)\rho+c\rho_{h}$. The scalar
field $c$ is assumed to evolve according to
\begin{equation}
\pr_{t} (\rho c) + \pr_{j}(u_{j}\rho c) - \frac{1}{Sc} \pr_{jj}c =0
\label{pascal_eq}
\end{equation}
where $Sc$ denotes the Schmidt number. The scalar field $c$ is convected by the flow
field $u_{j}$ but in turn affects the flow through the source terms in \ref{nseqs}. The initial state for $c$ is a step from a value 0
in the lower half to 1 in the upper half of the domain, representing a heavy fluid layer on top of a lighter one.

\section{Buoyancy effects in turbulent mixing}

To quantify buoyancy effects in turbulent mixing we consider the temporal mixing layer as documented in \cite{vremanjfm}. The flow
starts from a perturbed laminar profile. A fourth order accurate finite volume discretization was used. Rapid transition to
turbulence emerges, displaying helical pairings. At large buoyancy effects the separating interface becomes very distorted and protruding
heavy/light fluid `fingers' with a complex geometrical structure emerge, as shown in figure~\ref{snap_and_area_wrink_curv}(a).

\begin{figure}[hbt]

\centerline{
{\psfig{figure=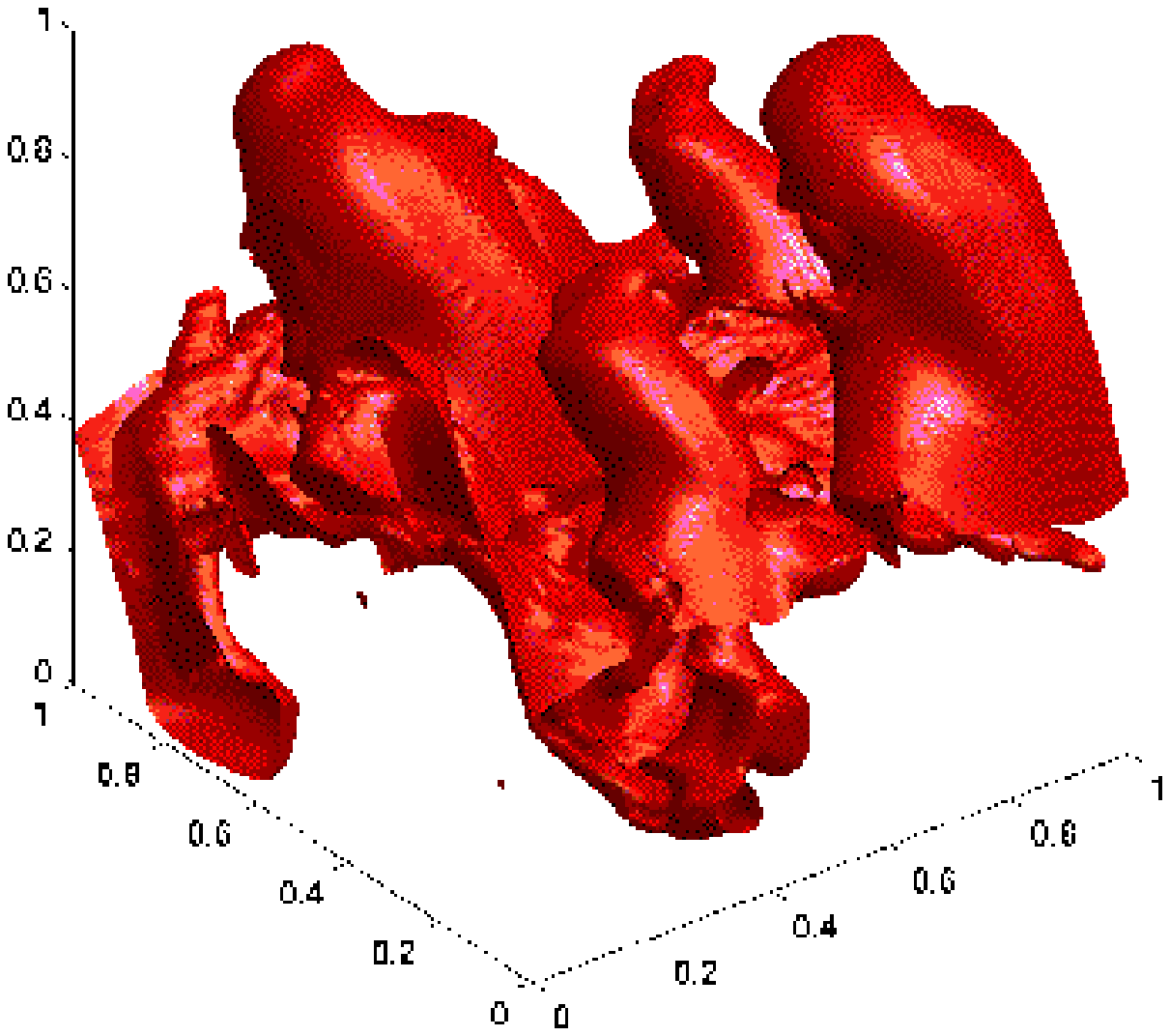,width=0.4\textwidth}} (a)
{\psfig{figure=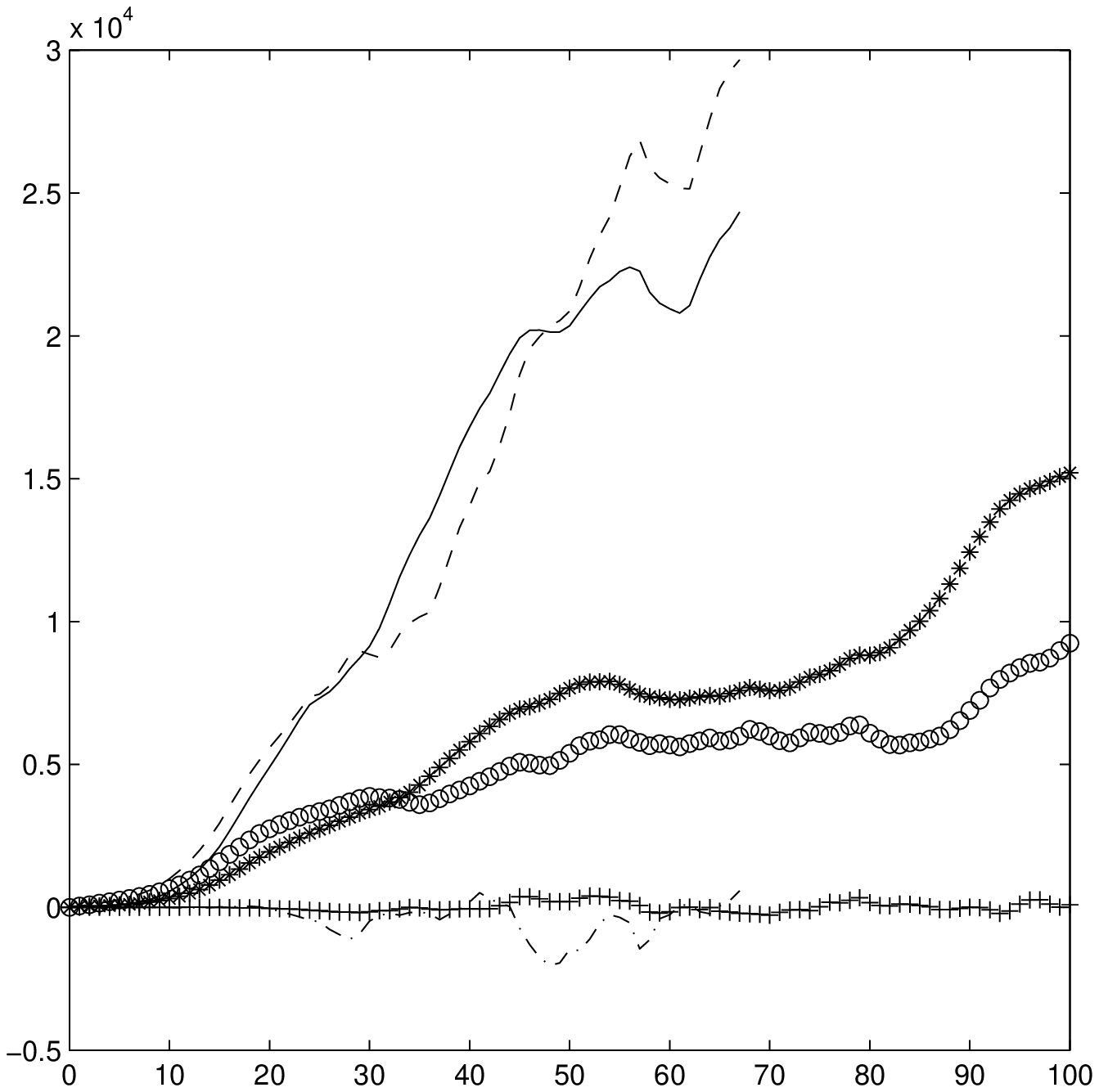,width=0.4\textwidth}} (b)
}

\vspace*{-0.35\textwidth}

\hspace*{0.4\textwidth} $A, C, W$

\vspace*{0.3\textwidth}

\hspace*{0.35\textwidth} $~$ \hspace*{0.4\textwidth} $t$

\vspace*{-3mm}

\caption{Separating interface at $t=40$ using $Fr=2$ (a); surface-area, wrinkling and
curvature (b) at $Fr=2$: $A(t)-A(0)$ (solid), $W$ (dashed), $C$ (dash-dotted) and $Fr=5$: $A(t)-A(0)$ (*), $W$ ($\circ$), $C$ (+).
Resolution of $128^{3}$ cells; $Sc=10$.}

\label{snap_and_area_wrink_curv}

\end{figure}

The geometrical properties $A$, $C$ and $W$ of the separating interface are compared in figure~\ref{snap_and_area_wrink_curv}(b) for
two different Froude numbers. The surface-area and wrinkling display approximately the same behavior, while the
average surface-curvature remains quite limited, indicating that regions with negative and positive local curvature are about equally
important in this flow.

\begin{figure}[hbt]

\centerline{
{\psfig{figure=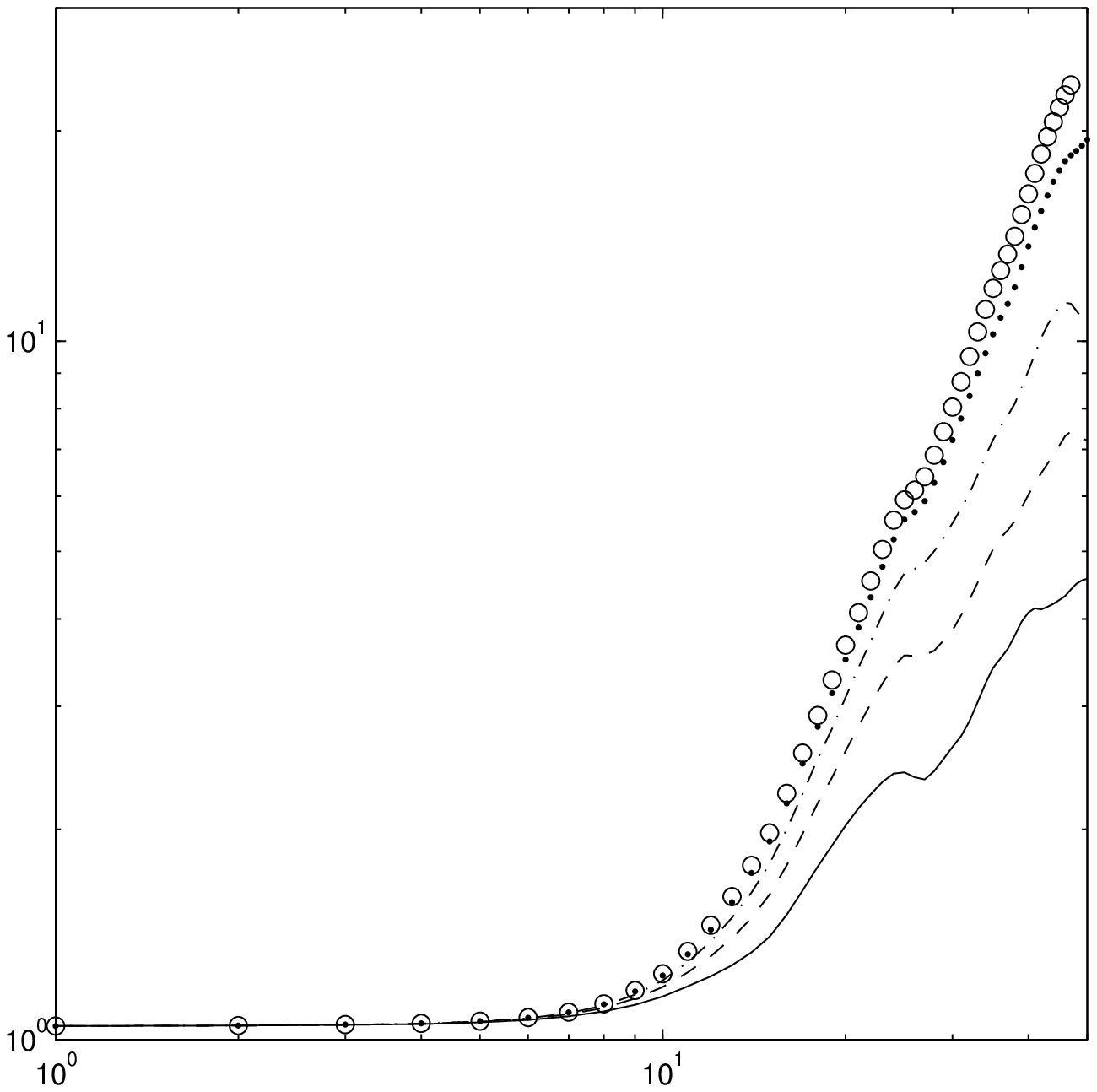,width=0.4\textwidth}} (a)
{\psfig{figure=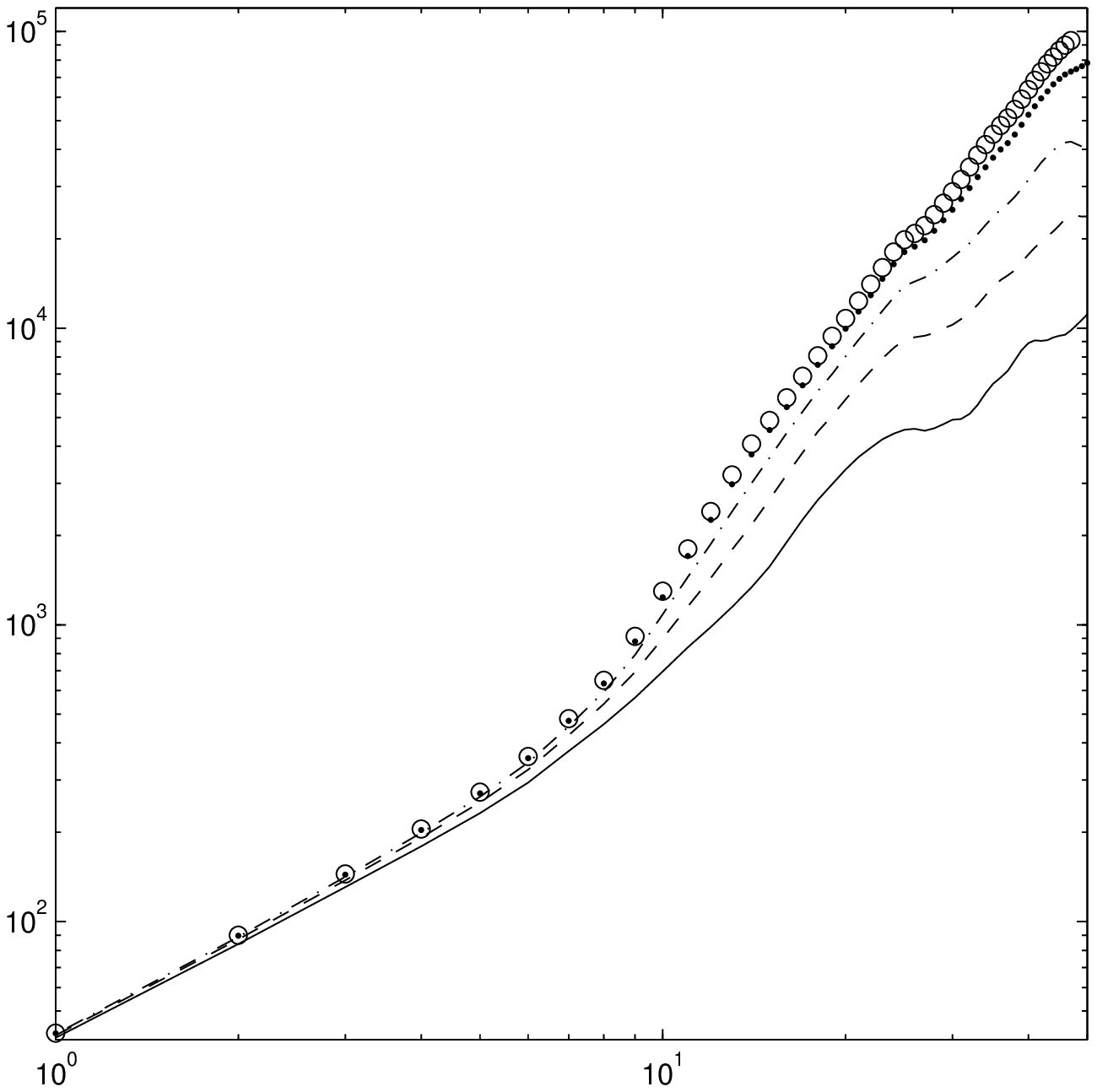,width=0.4\textwidth}} (b)
}

\vspace*{-0.35\textwidth}

\hspace*{0\textwidth} $\eta$ \hspace*{0.425\textwidth} $W$

\vspace*{0.3\textwidth}

\hspace*{0.35\textwidth} $t$ \hspace*{0.4\textwidth} $t$

\vspace*{-3mm}

\caption{Mixing-efficiency $\eta$ (a) and surface wrinkling (b) at $Fr=2$ and $Sc=5$ (solid), 10 (dashed), 20 (dash-dotted), 50 (dotted),
100 ($\circ$). Resolution: $128^{3}$.}

\label{schmidt_number_effects}
 
\end{figure}

In figure~\ref{schmidt_number_effects} we compare $\eta$ and $W$ for various $Sc$. An increase in $Sc$ reduces dissipation in
\ref{pascal_eq} and results in enhanced mixing. After a laminar development up to $t \approx 10$, a nearly algebraic growth
$\eta \sim t^{\beta}$ emerges with $\beta=\beta(Sc)$. The wrinkling also develops algebraically. Initially this development is
independent of $Sc$. Beyond $t \approx 10$ the algebraic growth-rate increases with $Sc$.

\begin{figure}[hbt]

\centerline{
{\psfig{figure=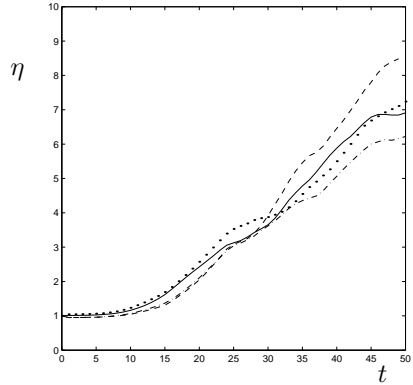,width=0.4\textwidth}}
}

\vspace*{-0.35\textwidth}

\hspace*{0.25\textwidth} $\eta$

\vspace*{0.3\textwidth}

\hspace*{0.65\textwidth} $t$

\vspace*{-3mm}

\caption{Mixing-efficiency $\eta$ at $Fr=2$ and $Sc=10$: DNS (solid), no-model $32^3$ (dashed), dynamic mixed model ($32^3$:
dash-dotted, $64^3$: dotted); $\Delta=L/16$.}

\label{les_of_buoyant_mixing}
 
\end{figure}

We also considered LES of this mixing flow. In figure~\ref{les_of_buoyant_mixing} we compare predictions from DNS and LES with the dynamic
mixed model. The filter-width $\Delta=L/16$ with $L$ the side of the flow domain. The mixing-efficiency is slightly under-predicted on
$32^{3}$ cells. The discrepancy with DNS is largely of numerical nature. By increasing the subgrid-resolution $r=\Delta/h$ from 2 on
$32^{3}$ to 4 on $64^{3}$ a strong improvement arises even though $\Delta$ was kept fixed. The remaining errors are more representative of
subgrid modeling deficiencies {\mbox{\cite{geurtsfroehlich2002}.}}

\vspace*{-2mm}

\section{Concluding remarks}
\label{concl}

Buoyancy can strongly increase turbulent mixing. A reasonable agreement between surface-area and wrinkling was observed. This may be
relevant, e.g., in combustion modeling to capture flame-speed and intensity. Prediction of mixing-efficiency using LES with dynamic models,
showed a large influence of spatial discretization errors. A subgrid-resolution $r=4$ is suggested.

\end{etcpaper}


\vspace*{-3mm}

\begin{thebibliography}{99}

\bibitem{haertel}
C. H\"artel, E. Meiburg, F. Necker. Analysis and direct
numerical simulation of the flow at a gravity-current head. Part 1:
Flow topology and front speed for slip and no-slip
boundaries. {\it J. Fluid Mech}~{{418}}: 189, 2000.

\bibitem{jot2001}
B.J. Geurts. Mixing efficiency in turbulent shear layers. {\it JOT}, {{2}}: 17, 2001.

\bibitem{geurtsfroehlich2002}
B.J. Geurts, J. Fr\"ohlich.
\newblock Numerical effects contaminating LES; a mixed story.
\newblock {\em Modern strategies for turbulent flow simulation}, Ed: B.J. Geurts, Edwards Publishing. 317-347, 2001.

\bibitem{vremanjfm} 
A.W. Vreman, B.J. Geurts, J.G.M. Kuerten.
\newblock  Large-eddy simulation of the turbulent mixing layer.
\newblock {\it J. Fluid Mech.} {339}: 357, 1997.

\end{thebibliography}
\end{document}